\documentclass[lettersize,journal]{IEEEtran}
\usepackage{amsmath,amsfonts}
\usepackage{algorithmic}
\usepackage{array}
\usepackage[caption=false,font=normalsize,labelfont=sf,textfont=sf]{subfig}
\usepackage{textcomp}
\usepackage{stfloats}
\usepackage{url}
\usepackage{verbatim}
\usepackage{graphicx}
\usepackage{algorithm}
\usepackage{tabularx}
\usepackage{pifont}
\usepackage{mathtools}
\usepackage{environ}
\usepackage{multirow}
\usepackage{booktabs}
\def\BibTeX{{\rm B\kern-.05em{\sc i\kern-.025em b}\kern-.08em
    T\kern-.1667em\lower.7ex\hbox{E}\kern-.125emX}}
\usepackage{balance}

\usepackage{xcolor}
\usepackage{colortbl}
\definecolor{myred}{RGB}{192, 0, 0}
\definecolor{myorange}{RGB}{217, 149, 44}
\definecolor{lightgraybg}{RGB}{240, 240, 240}

\definecolor{StepBlue}{RGB}{0, 102, 204} 
\definecolor{StepGold}{RGB}{204, 153, 0}

\usepackage{tcolorbox}
\definecolor{MyBoxBg}{RGB}{253, 252, 245}  
\definecolor{MyBoxFrame}{RGB}{100, 100, 100} 

\newcommand{\ignore}[1]{}

\newtcolorbox{MyQuestionBox}{
    colback=white,       
    colframe=black,      
    boxrule=0pt,         
    toprule=0.8pt,       
    bottomrule=0.8pt,    
    sharp corners,       
    left=2pt, right=2pt, 
    top=4pt, bottom=4pt, 
    before skip=10pt,    
    after skip=10pt,     
    fontupper=\itshape\small 
}

\newtcolorbox{MySoftBox}{
    colback=gray!5,      
    colframe=gray!40,    
    arc=3pt,             
    boxrule=0.5pt,       
    left=6pt, right=6pt, 
    top=5pt, bottom=5pt,
    fontupper=\itshape
}

\begin{document}
\title{From Retrieval to Reasoning: A Framework for Cyber Threat Intelligence NER with Explicit and Adaptive Instructions}

\author{Jiaren~Peng,
        Hongda~Sun,
        Xuan~Tian,
        Cheng~Huang,
        Zeqing~Li,
        and~Rui~Yan%
\thanks{J. Peng and H. Sun contributed equally to this work. (Corresponding author: Cheng Huang.)}%
\thanks{J. Peng, X. Tian, C. Huang, and Z. Li are with Sichuan University, Chengdu, China (e-mail: jiarenpeng666@gmail.com; xuantian@stu.scu.edu.cn; codesec@scu.edu.cn; alkarssec@gmail.com).}%
\thanks{H. Sun is with the Gaoling School of Artificial Intelligence, Renmin University of China, Beijing, China (e-mail: hongdasun985@gmail.com).}%
\thanks{R. Yan is with the School of Artificial Intelligence, Wuhan University, Wuhan, China, also with the Gaoling School of Artificial Intelligence, Renmin University of China, Beijing, China, and also with the Engineering Research Center of Next-Generation Intelligent Search and Recommendation, Ministry of Education, China (e-mail: rui.yan.pku@gmail.com).}%
}

\markboth{IEEE Transactions on , November~2025}%
{How to Use the IEEEtran \LaTeX \ Templates}

\maketitle

\begin{abstract}
The automation of Cyber Threat Intelligence (CTI) is central to proactive cyber defense, a process that heavily relies on Named Entity Recognition (NER) technology to extract critical entities from unstructured text. Currently, Large Language Models (LLMs) primarily address this task through the retrieval-based In-Context Learning (ICL) paradigm, guiding model predictions by retrieving similar examples. This paper conducts an in-depth analysis of this mainstream paradigm to uncover the driving factors behind its performance. Our analysis reveals a fundamental flaw: its success does not primarily stem from global semantic similarity but is highly dependent on the incidental overlap of entity types within the retrieved examples. This finding exposes the inherent limitation of its reliance on unreliable implicit induction.
To address this issue, we propose a novel paradigm that shifts from implicit induction to explicit instruction. We design TTPrompt, an instruction framework whose architecture directly maps the core concepts of CTI's Tactics, Techniques, and Procedures (TTPs). It formulates high-level task definitions as Tactics, guiding strategy as Techniques, and annotation guidelines as Procedures, thereby constructing a complete extraction pathway for the LLM from strategy to execution. Furthermore, to overcome the adaptability challenge of static guidelines, we introduce the Feedback-driven Instruction Refinement (FIR) mechanism. FIR enables the LLM to self-refine its guidelines set by learning from errors on a minimal amount of labeled data, thereby achieving rapid adaptation to the distinct annotation dialects of various datasets.
Experiments on five CTI NER benchmark datasets demonstrate that TTPrompt consistently surpasses retrieval-based baselines, but more importantly, its performance after instruction refinement on just 1\% of the training data~\footnote{5\% for CTINexus, i.e., 6 samples, due to its small size. } 
is comparable to that of models fine-tuned on the full dataset. For instance, on LADDER, its Micro F1 score of 71.96\% approaches the fine-tuned baseline. On the more complex CTINexus, its Macro F1 exceeds that of the fine-tuned ACLM model by 10.91\%.

\end{abstract}


\section{Introduction}
\label{sec:Introduction}

In today's rapidly evolving cyber threat landscape, the ability to swiftly extract actionable intelligence from unstructured text is paramount for building proactive cyber defense systems~\cite{xu2024intelex,hu2024llmtikg,guo2023framework,rahman2023attackers}. NER serves as a core technology for this goal, responsible for locating and classifying the atomic elements that constitute adversarial TTPs~\cite{cai2025knowledge}. However, the traditional fine-tuning paradigm, represented by models like BERT, struggles to keep pace with the fast-evolving threat environment due to its reliance on large-scale annotated data and limited generalization~\cite{cheng2025ctinexus}.

In recent years, leveraging retrieval-based In-Context Learning (RetICL)~\cite{luo2024RetICL} to unlock the few-shot capabilities of LLMs has become a mainstream paradigm for NER~\cite{cheng2025ctinexus,luo2025oneke,lekssays2025techniquerag}. However, an inherent disconnect exists between this paradigm's retrieval criteria and its learning process. Specifically, the retrieval process selects examples based solely on textual semantic similarity, yet each example presented to the LLM contains two sources of information that influence its decision: the contextual information carried by the text itself, and the entity annotation examples attached to it. As the retrieval criteria completely ignore the latter, a fundamental question arises: 
\begin{MyQuestionBox}
To what extent does an LLM's successful annotation depend on imitating similar text contexts, versus benefiting from the annotation examples that happen to be attached?
\end{MyQuestionBox}
We term this process of independently summarizing and inferring annotation rules from specific examples as implicit induction. Our experiments reveal that the success of RetICL is highly dependent on the incidental overlap of entity types in the retrieved samples, exposing its fundamental limitation of relying on unreliable implicit induction.

The aforementioned findings have motivated us to explore a new paradigm that shifts from implicit induction to explicit instruction. To this end, we have designed TTPrompt, an instruction hierarchy whose architecture directly maps the analytical framework used by CTI experts for TTPs. It treats the overall instruction as the Tactic, strategies as the Technique, and precise annotation guidelines as the Procedure, thereby constructing a complete deductive pathway for the LLM from strategy to execution. However, any static guidelines set faces an adaptability challenge, struggling to handle the unique annotation dialects of different datasets. To address this, we further introduce the FIR mechanism. FIR is an update process that conceptually mimics backpropagation, enabling the LLM to iteratively refine guidelines based on errors generated from a small amount of labeled data. This mechanism shifts the object of optimization from the model's internal parameters to the guidelines themselves, which serve as external inpuNOVEMBER ts.

Experiments show that our model consistently outperforms methods based on the RetICL paradigm across both micro and macro F1 scores on five distinct CTI NER datasets. On some datasets, its performance surpasses that of paradigms that fine-tune on the entire dataset. Furthermore, our quantitative analysis of the datasets' extraction difficulty indicates that our method demonstrates stronger extraction capabilities on more challenging datasets. The main contributions of this paper are as follows:

$\bullet$ We are the first to empirically demonstrate the underlying mechanism of RetICL in NER tasks, revealing its dependence on the incidental overlap of entity types and its limitations.

$\bullet$ We propose TTPrompt, a novel hierarchical instruction framework that mimics expert cognition to construct a deductive reasoning pathway for LLMs.

$\bullet$ We design FIR, a data-efficient mechanism that enables the annotation guideline to dynamically adapt to dataset-specific annotation dialects via instruction backpropagation.

$\bullet$ Experiments on 5 CTI benchmark datasets validate that FIR consistently outperforms sophisticated RetICL baselines and, on datasets with higher annotation difficulty, is competitive with mainstream models fine-tuned on the full dataset.

\section{Related Work}

\subsection{Cyber Threat Intelligence}
CTI is evidence-based knowledge about existing or emerging cyber threats that can be used to guide an organization's defensive decisions~\cite{rahman2023attackers,alam2023ladder}. CTI is crucial for proactive defense strategies. For example, it supports key activities like Threat Hunting, helping analysts to proactively search within their networks for potential attackers who have bypassed traditional defenses~\cite{maxam2024interview}. However, despite its immense value, a large volume of CTI remains scattered in the form of unstructured text across sources such as security reports and technical blogs. Therefore, how to automatically extract structured knowledge from these sources is one of the core challenges in the CTI domain today.

In the CTI automation pipeline, NER is a fundamental and critical step. Early NER methods primarily employed deep learning models such as BiLSTM-CRF and improved performance through techniques like joint extraction~\cite{guo2023framework}. Subsequently, fine-tuning methods based on PLM like BERT became mainstream, significantly boosting NER accuracy~\cite{alam2023ladder}. However, such fine-tuning methods, which rely on large-scale manual annotation, generally face a core bottleneck: they are not only costly to develop but also have poor scalability when faced with rapidly evolving new threats and entities. As a more flexible alternative, the ICL paradigm has emerged. It can perform high-precision NER tasks with only a few examples provided in a prompt, without any model fine-tuning~\cite{cheng2025ctinexus}.

\subsection{In-Context Learning}
The ICL ability exhibited by LLMs, which involves performing new tasks through a few demonstration examples, is a central topic in current research~\cite{dong2024survey}. Early research primarily focused on optimizing ICL performance, with a key direction being the improvement of example selection. Consequently, researchers proposed RetICL, whose core idea is to dynamically retrieve the most relevant examples for each input query from a large pool of examples~\cite{luo2024RetICL}. 
To achieve efficient retrieval, subsequent work further developed methods such as training dedicated retrievers using knowledge distillation, aiming to precisely acquire the highest quality examples~\cite{wang2024learning}.


As the question of \textbf{\textit{what examples to select?}} was studied in depth, the community shifted to a more fundamental question: \textbf{\textit{how do models learn from these examples?}} ICL has been interpreted as a form of implicit fine-tuning~\cite{dai2023can}, where the LLM is argued to act as a meta-optimizer in this process, capable of dynamically generating meta-gradients based on the examples to construct a model adapted to the current task. Another study, based on an information flow perspective, found that label words in the examples act as anchors: in the model's shallower layers, information first converges on them; subsequently, in the deeper layers, the model extracts the information from these anchors to make its final prediction~\cite{wang2023label}.
When research expanded to complex reasoning tasks, it was discovered that for the reasoning process, the logical structure of an example is more important than the correctness of its content. For instance, it has been demonstrated that a model can effectively learn a reasoning framework and patterns from a CoT that is logically coherent but contains factual errors, revealing that the model primarily learns a generalizable way of thinking~\cite{li2025llms}.


This insight naturally leads to a deeper question: If the core of what the model learns is the methodology rather than individual cases, is relying on implicit induction from scattered and unstable examples the best way to convey this methodology~\cite{zhou2023heuristics}? We argue that a more direct path is to provide this general framework to the model in the form of explicit instructions. To this end, the TTPrompt proposed constructs a complete deductive pathway that simulates expert cognition, from tactics to procedures. This represents a paradigm shift from implicit induction to explicit guidance.

\section{Retrieval-based ICL}
\label{sec:Retrieval-based ICL}
\subsection{Task Definition}
The core objective of general NER is to locate and classify entity mentions of predefined categories from unstructured text.
Formally, given an input text $D$ and a predefined set of entity types $Y=\{y_{1}, y_{2}, ..., y_{M}\}$, the NER task aims to extract a set of entity mentions $\mathcal{E}$ from $D$. Each entity $\mathcal{E}_{j} \in \mathcal{E}$ is a tuple $(\text{span}, y_{m})$, where $\text{span}$ represents the textual content of the entity in the original text, and $y_{m} \in Y$ is its corresponding entity type.

In the mainstream supervised learning paradigm, this task is typically solved by training a parameterized model $\mathcal{M}_{\theta}$. This model learns the mapping from the input text to the set of entities. Specifically, the optimization centers on the model parameters $\theta$. The training process seeks the optimal parameter set $\theta^*$ by minimizing a differentiable loss function $\mathcal{L}$ on an observable, labeled dataset $D_{u} = \{(D^{(j)}, \mathcal{E}^{(j)})\}$. This is generally achieved through a gradient-based backpropagation algorithm, and optimization objective can be formalized as:
\begin{equation}
\theta^* = \arg\min_{\theta} \mathcal{L}(D_{u}; \theta)
\label{for:optimization objective}
\end{equation}
where the loss function $\mathcal{L}$ typically uses Negative Log-Likelihood to measure the discrepancy between the model's predictions and the ground-truth labels:
\begin{equation}
\mathcal{L}(D_{u}; \theta) = - \sum_{(D^{(j)}, \mathcal{E}^{(j)}) \in D_{u}} \log P(\mathcal{E}^{(j)}|D^{(j)}; \theta)
\end{equation}

\subsection{Retrieve-and-Prompt}
To better investigate this problem, inspired by CTINEXUS*~\cite{cheng2025ctinexus}~\footnote{To distinguish between datasets and models with the same name, this paper uses LADDER*~\cite{alam2023ladder} and CTINexus* to denote the models.}, we designed a method based on the RetICL paradigm as the subject of our study. The core idea is to dynamically retrieve the most relevant examples for each document to be processed and construct a prompt containing instructions, examples, and the query to guide an LLM in completing the NER task. For a comprehensive description of the specific implementation and setup, please refer to Appendix~\ref{sec:kNN-based}~\footnote{\textbf{Statistical Significance Declaration:} To ensure the reliability of our conclusions, for all comparative experiments reported in this paper, we performed a t-test comparing the results against the second-best baseline, confirming a statistical significance level of $p < 0.05$.}.

\subsection{Intrinsic Mechanisms of RetICL}
This section aims to dissect the intrinsic mechanisms and core limitations of the existing RetICL paradigm in CTI NER tasks through a series of comparative experiments and to 
empirically answer the fundamental research question raised in Section~\ref{sec:Introduction}. 

\subsubsection{\textbf{Baseline Performance of Semantic-kNN}}

\begin{figure}
    \centering
    \includegraphics[width=1\linewidth]{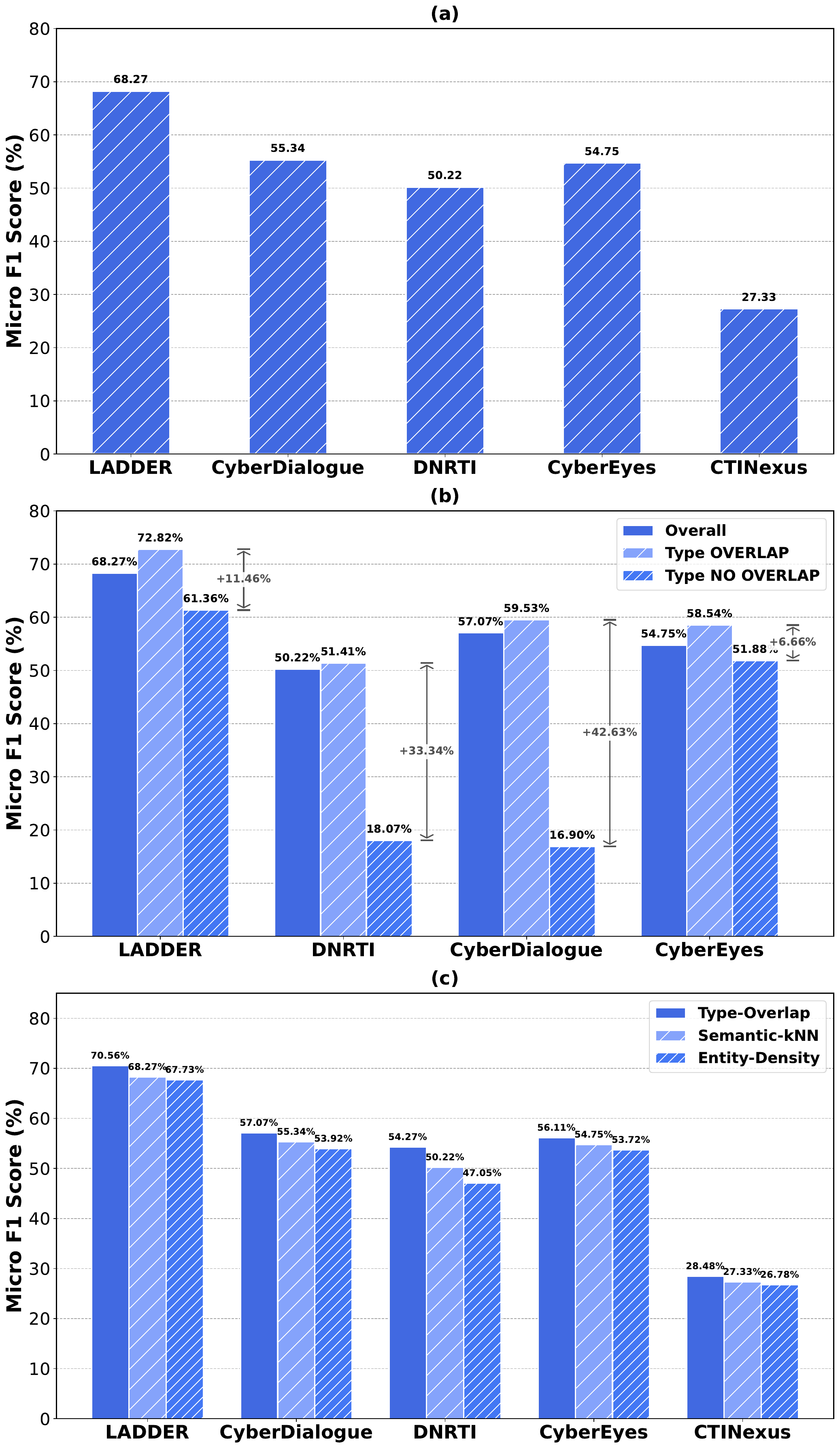}
    \caption{Empirical analysis of the RetICL mechanism. \textbf{(a)} Baseline performance of the Semantic-kNN method across five CTI NER benchmarks. \textbf{(b)} Impact of entity type overlap on extraction performance. The significant gap between "Type OVERLAP" and "Type NO OVERLAP" reveals the model's heavy reliance on incidental type matching. Note that CTINexus is omitted here as all its cases fall under Type OVERLAP. \textbf{(c)} Performance comparison of three distinct retrieval paradigms. The superiority of the Type-Overlap paradigm over Semantic-kNN and Entity-Density further verifies that explicit type guidance is more effective than implicit semantic induction.}
    \label{fig:Semantic-kNN_Base}
\end{figure}


We first evaluated the performance of the current mainstream context learning method based on semantic similarity retrieval, Semantic-kNN, on five CTI NER benchmark datasets. The experimental results are shown in Figure~\ref{fig:Semantic-kNN_Base}\textbf{(a)}. Semantic-kNN as a powerful baseline method achieves good base performance across the datasets, reaching an F1 Score of 68.27\% on LADDER for example. \looseness=-1

\subsubsection{\textbf{The Critical Role of Entity Type Overlap}}
To deeply analyze the intrinsic mechanism behind the baseline performance of the Semantic-kNN method, we deconstructed its core process. The method selects demonstration examples based on global semantic similarity, meaning the retrieval process is driven entirely by Textual Content. However, once an example is selected and placed in the prompt, the LLM is exposed to an information body with two dimensions: one is the textual content of the example itself, and the other is the entity annotation exemplars attached to that text. A core unknown is which of these two contributes more critically to the model's performance. \looseness=-1



To decouple and quantify the influence of these two information sources, we designed an in-depth analysis experiment. The test results of Semantic-kNN on each dataset were divided into two subsets for independent evaluation based on whether entity type overlap exists between the k retrieved examples and the query text's ground truth. The results are shown in Figure~\ref{fig:Semantic-kNN_Base}\textbf{(b)}. \looseness=-1

The experimental results indicate: 
(1) In cases where there is entity type overlap between the retrieved examples and the query text, Type OVERLAP, the model's performance is far higher than the overall average, reaching 72.82\% on LADDER for instance.
(2) Conversely, when no entity type overlap exists between the examples and the query text, Type NO OVERLAP, the model's performance suffers a precipitous decline. On DNRTI~\cite{wang2020dnrti} and CyberDialogue~\cite{yang2023cyberdialogue}, for example, the scores dropped by 33.34\% and 42.63\% respectively.

This performance disparity shows that the success of Semantic-kNN is largely not because it finds the most semantically similar texts, but because it luckily finds examples containing the target entity types. 
This observation provides a clear answer to our guiding question in Section~\ref{sec:Introduction}: for an LLM, 
learning the annotation patterns of specific entity types is more critical than understanding broad textual information.


\subsubsection{\textbf{From Implicit to Explicit}}
Based on the above hypothesis, we further designed an experiment to verify the dominant role of entity type overlap, and we implemented two additional retrieval paradigms.

These two paradigms are: 
\textbf{(1) Type-Overlap:} This paradigm disregards semantics and directly selects the k examples with the most entity type overlap with the query text. It must be emphasized that Type-Overlap is not a practically applicable paradigm but an Oracle Setting. It operates on the unrealistic premise that we know the ground truth entity types in the query text beforehand, as detailed in Appendix~\ref{sec:Type-Overlap}.
\textbf{(2) Entity-Density:} This paradigm completely ignores the query text and always selects the k examples from the dataset that contain the highest number of entities themselves, as detailed in Appendix~\ref{sec:Entity-Density}.

From Figure~\ref{fig:Semantic-kNN_Base}\textbf{(c)}, we can draw two conclusions:
\textbf{(1). Explicit induction is superior to implicit induction.} The Type-Overlap paradigm outperforms Semantic-kNN on almost all datasets; for instance, performance on LADDER increased from 68.27\% to 70.56\%. This strongly confirms our core hypothesis: model performance improves significantly when the retrieval paradigm's objective shifts from vague semantic similarity to explicit type matching.
\textbf{(2). Simple entity piling is counterproductive.} Meanwhile, the Entity-Density paradigm performs the worst, far below even Semantic-kNN. This result indicates that merely providing the LLM with dense samples full of entities is ineffective. The model struggles to learn effective annotation rules for entity types from such untargeted implicit information and instead becomes confused by overly complex patterns.

\begin{figure*}[!h]
    \centering
    \includegraphics[width=1\linewidth]{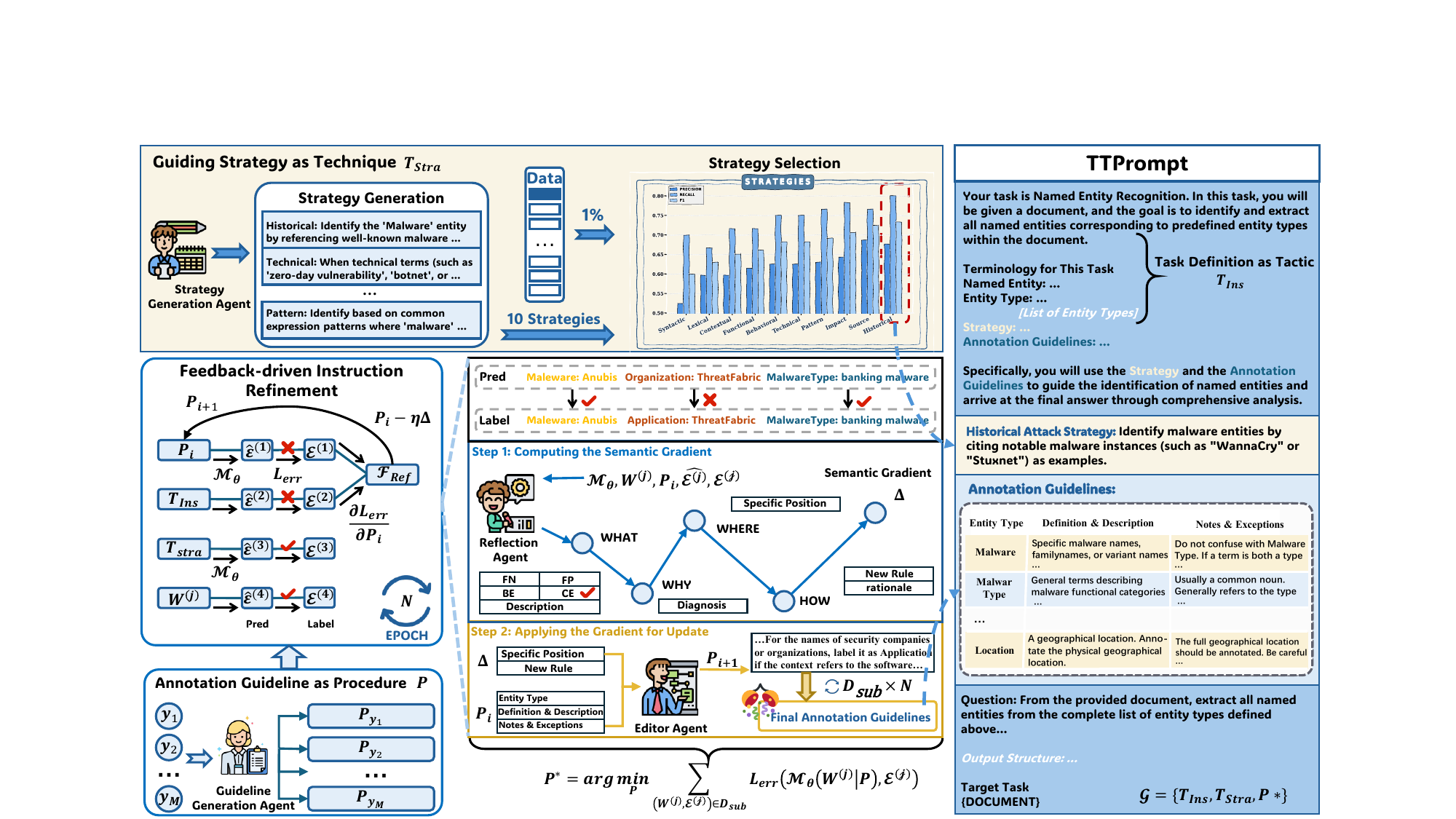}
    \caption{This is a system overview of our TTPrompt framework. The process involves generating initial annotation Guidelines, selecting the optimal strategy, and iteratively refining these guidelines through the Feedback-driven Instruction Refinement mechanism. The Reflection Agent and Editor Agent are used to implement the core refinement loop; they calculate a semantic gradient from prediction errors to progressively optimize the final instruction set used for inference.}
    \label{fig:main}
\end{figure*}

\section{TTPrompt}
The empirical analysis in Section~\ref{sec:Retrieval-based ICL} revealed a fundamental limitation of the RetICL paradigm on NER tasks: its performance is overly dependent on the incidental overlap of entity types in the demonstration examples. At a deeper level, this method based on implicit induction not only exhibits inherent instability but also fails to effectively handle complex and variable CTI scenarios. These findings collectively point to an urgent need for an explicit method to guide LLMs.

In response to these challenges, as shown in Figure~\ref{fig:main}, we propose the novel TTPrompt framework. This framework consists of two parts. First, we construct a Hierarchical Instruction Framework designed to replace the reliance on implicit induction with explicit expert knowledge, directly teaching the LLM how to perform annotation. Second, to overcome the inherent limitations of static instructions in adapting to the annotation guidelines of different datasets, we introduce FIR, which enables instructions to evolve adaptively through iteration. \looseness=-1

\subsection{Hierarchical Instruction Framework}
To address the core need of letting the model know how to annotate, our designed instruction set is not a flat list of guidelines but a structured object that mimics the cognitive hierarchy of human experts.

\subsubsection{\textbf{Task Definition as Tactic}}
The Tactic layer is the highest level of the instruction framework. This layer corresponds to the instruction part $T_{Ins}$ in TTPrompt. The instruction specifies the task the LLM needs to complete and the desired output format.

\subsubsection{\textbf{Guiding Strategy as Technique}}
The guiding principles layer is the cognitive strategy layer in the TTPrompt framework, closely associated with the concept of Techniques in the TTPs framework. These strategies guide the model's thinking path at a macro level. This paper constructs the instructions for this layer through a two-stage automated process.

\textbf{Strategy Generation.}
To ensure the diversity and high quality of the guiding principles, we first construct a strategy generation agent. We abstract the function of this agent as a generation function $\mathcal{F}_{StratGen}$, which takes the desired number of strategies $n=10$ as input and outputs an initial set $S$ containing several different strategies. This process can be formalized as:
\begin{equation}
\mathcal{S}=\mathcal{F}_{StratGen}(n)=\{s_{1},s_{2},...,s_{n}\}
\end{equation}

\textbf{Strategy Selection.}
Since not all generated strategies are equally applicable to the target NER task, we introduce a strategy selection step to filter out low-quality ones. Specifically, we independently evaluate the performance of each strategy $s_{j}$ in the set $S$ on a very small training subset $D_{sub}$ with its corresponding ground truth labels $\mathcal{E}_{sub}$. The size of this subset is 1\% of $D_{u}$, consistent with the one-shot definition.
We abstract this evaluation process into a scoring function $\mathcal{F}_{Eval}$, which calculates a performance score $\sigma_{j}$ for each strategy $s_{j}$. This process can be formalized as:
\begin{equation}
\sigma_{j}=\mathcal{F}_{Eval}(\mathcal{M}_{\theta}(D_{sub}|T_{ins},s_{j}),\mathcal{E}_{sub})
\end{equation}
Here, $\mathcal{M}_{\theta}$ represents the language model. Finally, we select the strategy that maximizes this score and denote it as $T_{Stra}$.
\begin{equation}
T_{Stra}=\arg\max_{s_{j}\in\mathcal{S}}\sigma_{j}
\end{equation}
Through this strategy selection step, we can effectively exclude low-quality strategies.

\subsubsection{\textbf{Annotation Guideline as Procedure}}
The Annotation Guideline $P$ component is responsible for providing the operational Execution Protocol for this process. It translates the macro-level thinking initiated by $T_{Stra}$ into precise annotation actions. This layer of instruction conceptually corresponds to the specific Procedures that CTI experts follow in their analysis. \looseness=-1

Similar to the strategy generation process, we also construct a guideline generation agent. Its task is to generate a set of structured annotation guidelines for a predefined set of entity types $\mathcal{Y}=\{y_{1},y_{2},...,y_{M}\}$.

For any entity type $y_{m}\in\mathcal{Y}$, the agent generates specifications covering the following aspects:
(1) Definition and Description: Clearly specifies the start and end criteria for entity text spans.
(2) Notes and Exceptions: Explicitly demonstrates how to handle easily confused entity types.

We abstract the generation process of this agent into a function denoted as $\mathcal{F}_{SpecGen}$. This function takes the set of entity types as input and outputs a set of structured specifications for them, which we denote as P. The process can be formalized as:
\begin{equation}
P=\bigcup_{y_{m}\in\mathcal{Y}}P_{y_{m}}=\bigcup_{y_{m}\in\mathcal{Y}}\mathcal{F}_{SpecGen}(y_{m})
\end{equation}
In summary, the complete initial instruction set of our methodology is denoted as $\mathcal{G}=\{T_{Ins},T_{Stra},P\}$.

\subsection{Feedback-driven Instruction Refinement}
Although the hierarchical instruction framework solves the problem of explicit rule execution, any static annotation guidelines $P$ faces a key challenge: annotation discrepancies across datasets. To solve this, we designed FIR, an iterative optimization process that conceptually mimics gradient descent. However, it shifts the optimization from a continuous parameter space to a discrete instruction space.

In the traditional paradigm, the optimization objective is shown in Equation~\ref{for:optimization objective}. In contrast, FIR operates on a model $\mathcal{M}_{\theta}$ with frozen parameters, and its optimization target is the annotation guideline $P$. The process is driven by a binary error signal $L_{err}$, defined as:
\begin{equation}
L_{err}(\hat{\mathcal{E}}^{(j)},\mathcal{E}^{(j)})=\mathbb{I}(\hat{\mathcal{E}}^{(j)}\ne\mathcal{E}^{(j)})
\end{equation}
The FIR framework iterates for a number of epochs over a very small training subset $D_{sub}$. In the $i$-th iteration, for any sample $(W^{(j)},\mathcal{E}^{(j)})$ in $D_{sub}$, the following two stages are executed.

\subsubsection{\textbf{Instruction Forward Pass}}
The goal of this stage is to generate predictions. The parameter-frozen model $\mathcal{M}_{\theta}$ processes the input $W^{(j)}$ under the guidance of the current instructions $\mathcal{G}$ to output a predicted set of entities $\hat{\mathcal{E}}^{(j)}$. This forward pass can be formalized as:
\begin{equation}
\hat{\mathcal{E}}^{(j)}=\mathcal{M}_{\theta}(W^{(j)}|\mathcal{G})
\end{equation}

\subsubsection{\textbf{Instruction Backpropagation}}
This stage is activated only when there is a discrepancy between the prediction $\hat{\mathcal{E}}^{(j)}$ and the ground truth label $\mathcal{E}^{(j)}$, meaning $L_{err}=1$. The update target of this process is the instruction $P_{i}$ itself rather than the model parameters, accomplished through the following two steps. \looseness=-1

\textbf{\textcolor{StepBlue}{Step 1: Computing the Semantic Gradient.}}
The core of this step is to compute a gradient to guide the instruction optimization. Unlike the numerical gradient in traditional backpropagation that guides model parameter updates, we introduce a semantic gradient, denoted as $\Delta$. It is important to note that this is not a strict mathematical concept, but a conceptual analogy to the gradient descent optimization process. \looseness=-1

This semantic gradient is generated by a reflector agent. The agent takes a tuple $(W^{(j)},P_{i},\hat{\mathcal{E}}^{(j)},\mathcal{E}^{(j)})$ as input and performs a root cause analysis of the error through a structured Chain-of-Thought (CoT)~\cite{wei2022chain}. This process follows a What-Why-Where-How analytical framework:

\textbf{What}: First, the agent classifies the error according to four predefined types: False Negative (FN), False Positive (FP), Boundary Error (BE), or Classification Error (CE), and provides a description of the error.
\textbf{Why}: Second, it diagnoses the root cause of the problem, analyzing which rule in the current guideline led to the error.
\textbf{Where}: Next, it precisely locates the specific position in the guideline that needs to be modified.
\textbf{How}: Finally, it proposes how to correct it by generating a new rule text and explains the rationale for the modification. \looseness=-1

The set of revised instructions output by the agent constitutes the Semantic Gradient. Unlike a numerical gradient that only provides direction and magnitude, $\Delta$ not only precisely indicates the direction and extent of the optimization but also includes complete diagnostic information for the correction. This process can be abstracted as a function $\mathcal{F}_{Ref}()$, conceptually equivalent to taking the partial derivative with respect to the discrete instruction space:
\begin{equation}
\Delta=\mathcal{F}_{Ref}(\mathcal{M}_{\theta},W^{(j)},P_{i},\hat{\mathcal{E}}^{(j)},\mathcal{E}^{(j)})\equiv\frac{\partial L_{err}}{\partial P_{i}}
\end{equation}

\textbf{\textcolor{StepGold}{Step 2: Applying the Gradient for Update.}}
In traditional gradient descent, an optimizer such as Adam~\cite{loshchilov2017adam} performs the actual parameter update based on the computed gradient and learning rate: $\theta_{t+1}=\theta_{t}-\eta\nabla_{\theta}L$. In the FIR framework, this step is simulated by an editor agent. It receives the current instruction $P_i$ and the computed Semantic Gradient $\Delta$. By applying the modifications defined in $\Delta$ to $P_i$, it generates a new version of the instruction, $P_{i+1}$. This controlled integration process is conceptually equivalent to the parameter update performed by a traditional optimizer when applying a gradient. This process can be abstracted as a function $\mathcal{F}_{Upd}()$:
\begin{equation}
P_{i+1}=\mathcal{F}_{Upd}(P_{i},\Delta)=P_{i}-\eta\Delta
\end{equation}
The new guideline $P_{i+1}$ is immediately used to process subsequent samples, thus achieving continuous evolution of the instructions. The above process can be summarized as finding a final instruction set $P^{*}$ that minimizes the cumulative error signal by performing iterative semantic gradient descent on $D_{sub}$:
\begin{equation}
P^{*}= \arg\min_{P}\sum_{(W^{(j)},\mathcal{E}^{(j)})\in D_{sub}}L_{err}(\mathcal{M}_{\theta}(W^{(j)}|P),\mathcal{E}^{(j)})
\end{equation}

\section{Experiment Setup}

\subsection{Dataset and Evaluation}
To conduct a comprehensive, multi-dimensional evaluation of our proposed framework, this study selects five representative CTI NER benchmark datasets: LADDER~\cite{alam2023ladder}, DNRTI~\cite{wang2020dnrti}, CTINexus~\cite{cheng2025ctinexus}, CyberEyes~\cite{fang2021cybereyes} and CyberDialogue~\cite{yang2023cyberdialogue}. Detailed statistics for each dataset can be found in Appendix~\ref{sec:Datasets}. Additionally, we have designed a multi-dimensional metric for evaluating the NER difficulty of datasets, detailed in Appendix~\ref{sec:Difficulty}.

Following previous work~\cite{alam2023ladder,ghosh2023aclm}, we adopt Macro-averaged F1 (Macro-F1) and Micro-averaged F1 (Micro-F1)
as the primary evaluation metrics. For implementation details, please see the Appendix~\ref{sec:Implementation}.

\subsection{Baselines}

To fully evaluate the effectiveness of our method, we have selected four representative baseline models: \textbf{LADDER*}~\cite{alam2023ladder}, which embodies the traditional fine-tuning paradigm in the CTI domain; \textbf{ACLM}~\cite{ghosh2023aclm}, representing data augmentation techniques for complex and low resource fine-tuning NER environments; \textbf{OneKE}~\cite{luo2025oneke}, a multi-agent framework based on LLMs; and \textbf{CTINEXUS*}~\cite{cheng2025ctinexus}, an ICL method based on semantic similarity retrieval. Detailed descriptions of each baseline are provided in Appendix~\ref{sec:baseline_details}.

\begin{table*}[!htbp]
\setlength{\abovecaptionskip}{0.1cm}
\setlength{\belowcaptionskip}{-0.2cm}
    \centering
    \caption{Performance comparison of different models across five CTI NER datasets. The highest score among fine-tuning models is colored \textcolor{myred}{red}, and the highest among LLM ICL models is colored \textcolor{myorange}{orange}.}
    \label{tab:main}
    \begin{tabular}{l|cc|cc|cc|cc|cc}
        \hline
        \multirow{2}{*}{\textbf{Model}} & \multicolumn{2}{c|}{\textbf{LADDER}} & \multicolumn{2}{c|}{\textbf{CyberEyes}} & \multicolumn{2}{c|}{\textbf{CyberDialogue}} & \multicolumn{2}{c|}{\textbf{DNRTI}} & \multicolumn{2}{c}{\textbf{CTINexus}} \\
        \cline{2-11}
        ~ & Macro & Micro & Macro & Micro & Macro & Micro & Macro & Micro & Macro & Micro \\
        \hline \hline
        LADDER*    & \textcolor{myred}{69.56\%} & \textcolor{myred}{75.91\%} & 89.35\% & 87.50\% & 60.36\% & 78.86\% & 78.54\% & 78.38\% & 0.00\% & 0.00\% \\
        ACLM      & 67.49\% & 74.45\% & \textcolor{myred}{91.04\%} & \textcolor{myred}{88.86\%} & \textcolor{myred}{78.78\%} & \textcolor{myred}{87.46\%} & \textcolor{myred}{83.37\%} & \textcolor{myred}{82.77\%} & \textcolor{myred}{10.88\%} & \textcolor{myred}{25.53\%} \\ \hline
        OneKE     & 44.16\% & 45.85\% & 53.13\% & 45.06\% & 27.04\% & 35.40\% & 21.57\% & 33.22\% & 0.00\% & 0.00\% \\
        CTINEXUS*  & 62.67\% & 68.27\% & 62.69\% & 54.75\% & 41.22\% & 57.07\% & 47.62\% & 53.50\% & 17.89\% & 27.33\% \\
        \rowcolor{lightgraybg}
        \textbf{TTPrompt} & \textcolor{myorange}{67.92\%} & \textcolor{myorange}{71.96\%} & \textcolor{myorange}{64.55\%} & \textcolor{myorange}{56.52\%} & \textcolor{myorange}{46.92\%} & \textcolor{myorange}{58.05\%} & \textcolor{myorange}{48.17\%} & \textcolor{myorange}{61.62\%} & \textcolor{myorange}{21.79\%} & \textcolor{myorange}{31.41\%} \\
        \hline
    \end{tabular}
\end{table*}

\begin{table*}[!htbp]
\setlength{\abovecaptionskip}{0.1cm}
\setlength{\belowcaptionskip}{-0.2cm}
    \centering
    \caption{Ablation study for TTPrompt. For this table and all subsequent tables, the highest score in each column is highlighted in \textcolor{myred}{red} and the second-highest in \textcolor{myorange}{orange}, unless otherwise specified.}
    \label{tab:ablation_study}
    \begin{tabular}{l|cc|cc|cc|cc|cc}
        \hline
        \multirow{2}{*}{\textbf{Configuration}} & \multicolumn{2}{c|}{\textbf{LADDER}} & \multicolumn{2}{c|}{\textbf{CyberEyes}} & \multicolumn{2}{c|}{\textbf{CyberDialogue}} & \multicolumn{2}{c|}{\textbf{DNRTI}} & \multicolumn{2}{c}{\textbf{CTINexus}} \\
        \cline{2-11}
        ~ & Macro  & Micro  & Macro  & Micro  & Macro  & Micro  & Macro  & Micro  & Macro  & Micro  \\
        \hline \hline
        Base & 61.34\% & 61.63\% & 57.35\% & 44.71\% & 32.11\% & 39.14\% & 41.98\% & 49.11\% & 17.76\% & 27.49\% \\
        + $T_{Sta}$ & \textcolor{myorange}{64.32\%} & \textcolor{myorange}{68.18\%} & \textcolor{myorange}{60.59\%} & \textcolor{myorange}{50.67\%} & 32.96\% & 43.88\% & \textcolor{myorange}{46.13\%} & 53.79\% & \textcolor{myorange}{20.63\%} & \textcolor{myorange}{30.34\%} \\
        + $P$  & 62.12\% & 65.38\% & 57.03\% & 46.08\% & \textcolor{myorange}{37.25\%} & \textcolor{myorange}{47.89\%} & 43.50\% & \textcolor{myorange}{54.99\%} & 19.09\% & 27.44\% \\
        \hline
        \rowcolor{lightgraybg}
        \textbf{TTPrompt} & \textcolor{myred}{67.92\%} & \textcolor{myred}{71.96\%} & \textcolor{myred}{64.55\%} & \textcolor{myred}{56.52\%} & \textcolor{myred}{46.92\%} & \textcolor{myred}{58.05\%} & \textcolor{myred}{48.17\%} & \textcolor{myred}{61.62\%} & \textcolor{myred}{21.79\%} & \textcolor{myred}{31.41\%} \\
        \hline
    \end{tabular}
\end{table*}

\section{Experimental Results}
\begin{enumerate}
    \item[RQ1:] How does the overall performance of TTPrompt compare against mainstream baseline methods?
    \item[RQ2:] What are the contributions of the core components within TTPrompt to the final performance?
    \item[RQ3:] How does the extraction difficulty of different datasets affect methods from different paradigms?
    \item[RQ4:] Does the quality of guiding strategies depend on the model that generates them?
    \item[RQ5:] Can annotation guidelines be transferred from a teacher model to a student model?
    \item[RQ6:] What is the sensitivity of FIR to the number of iteration epochs and the amount of data?
\end{enumerate}

\subsection{RQ1: Main Result}

This section will analyze in detail the experimental results of TTPrompt and the baselines on the five datasets.

\textbf{TTPrompt vs. CTINEXUS*.}
As shown in Table~\ref{tab:main}, TTPrompt outperforms CTINEXUS* on all five datasets. For example, it leads by 5.7\% in Macro F1 on CyberDialogue and by 8.12\% in Micro F1 on DNRTI. This result strongly validates our core hypothesis: deductive reasoning based on explicit instructions is more reliable than RetICL methods that rely on implicit induction.

\textbf{TTPrompt vs. LADDER* \& ACLM.}
LADDER* and ACLM, as strong fine-tuning baselines, exhibit extremely competitive performance on some datasets. On the LADDER dataset, the LADDER* model, which was specifically optimized for that domain, achieved the highest Micro F1 score (75.91\%). However, it is noteworthy that our TTPrompt, using only 1\% of the data, can achieve performance comparable to these strong baselines on datasets like LADDER. Particularly on the CTINexus dataset, TTPrompt surpassed the traditional fine-tuning models.

\textbf{TTPrompt vs. OneKE.}
The experimental results show that OneKE's performance is significantly lower than other baselines across all datasets. We speculate this is mainly because a general-purpose framework struggles to handle the highly specialized nature of the CTI domain. Furthermore, OneKE's performance heavily relies on its internal Case Repository for multi-turn dialogue extraction, but a generic knowledge base not adapted for the CTI domain can hardly provide effective guidance for specialized extraction. In contrast, TTPrompt more robustly handles these challenges through its domain-customized explicit instructions and the FIR mechanism.

\subsection{RQ2: Ablation Study}

To quantify the contribution of each core component in the framework, we designed the ablation study in this section. We defined four variants: (1) Base, containing only the task instruction $T_{Ins}$; (2) + $T_{Sta}$, which adds the guiding strategy to the base; (3) + $P$, which adds the initially generated annotation guideline to the + $T_{Sta}$ variant; and (4) TTPrompt, which adds FIR to the + $P$ variant.

As shown in Table~\ref{tab:ablation_study}, the results reveal the interplay between the components. First, comparing base with + $T_{Sta}$, the latter achieves significant performance improvements on all datasets, with the Micro F1 score on the LADDER dataset increasing by 6.55\%. This indicates that the guiding strategy provides effective cognitive guidance. Second, a noteworthy phenomenon is that after adding the unrefined annotation guideline, the performance of + $P$ actually decreases; for example, its Micro F1 score on the LADDER dataset dropped by 2.8\%. This suggests that a generic set of guidelines not adapted to a specific dataset can have a detrimental effect on performance. Finally, comparing TTPrompt with + $P$, the introduction of the FIR mechanism not only compensated for the previous performance loss but also elevated all metrics to their highest levels. To provide a more intuitive demonstration of how FIR works, we present a detailed case study in Table~\ref{tab:case_study}.\looseness=-1


\begin{table*}[!h]
\setlength{\abovecaptionskip}{0.1cm}
\setlength{\belowcaptionskip}{-0.2cm}
    \centering
    \caption{Performance of baseline models fine-tuned on 1\% of the training data.}
    \label{tab:low_sft}
    \begin{tabular}{l|cc|cc|cc|cc|cc}
        \hline
        \multirow{2}{*}{\textbf{Model}} & \multicolumn{2}{c|}{\textbf{LADDER}} & \multicolumn{2}{c|}{\textbf{CyberEyes}} & \multicolumn{2}{c|}{\textbf{CyberDialogue}} & \multicolumn{2}{c|}{\textbf{DNRTI}} & \multicolumn{2}{c}{\textbf{CTINexus}} \\
        \cline{2-11}
        ~ & Macro  & Micro  & Macro  & Micro  & Macro  & Micro  & Macro  & Micro  & Macro  & Micro  \\
        \hline \hline
        LADDER* & 0.00\% & 0.00\% & 0.00\% & 0.00\% & 0.00\% & 0.00\% & 0.00\% & 0.00\% & 0.00\% & 0.00\% \\
        ACLM   & \textcolor{myred}{7.30\%} & \textcolor{myred}{11.23\%} & \textcolor{myred}{35.24\%} & \textcolor{myred}{43.59\%} & \textcolor{myred}{10.57\%} & \textcolor{myred}{19.42\%} & 0.00\% & 0.00\% & 0.00\% & 0.00\% \\
        \hline
    \end{tabular}
\end{table*}

\begin{figure*}[h]
    \centering
    \includegraphics[width=1\linewidth]{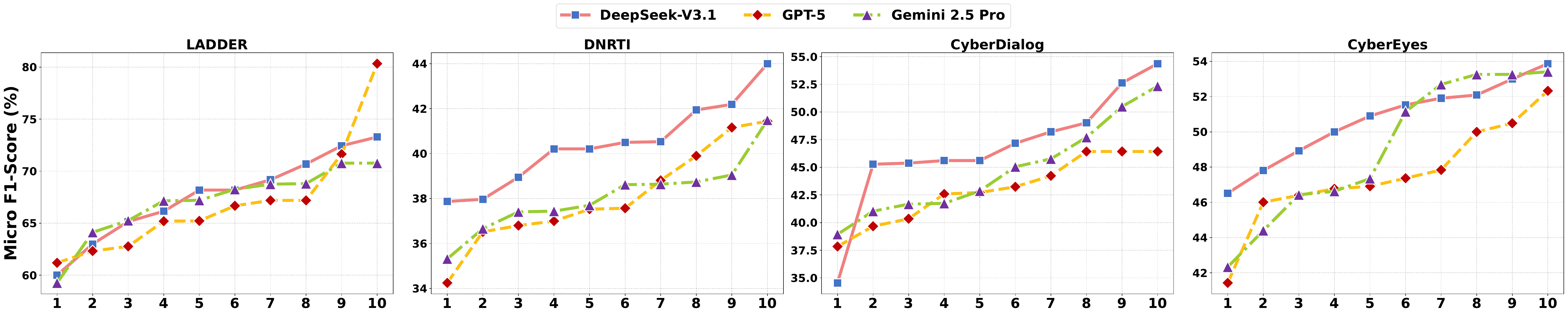}
    \caption{Performance comparison of strategies generated by three different LLMs.}
    \label{fig:Strategies}
\end{figure*}

\subsection{RQ3: Impact of Dataset Difficulty}

From Table~\ref{tab:main}, we can see that on datasets like LADDER, TTPrompt's performance is inferior to fully fine-tuned models. However, on the CTINexus dataset, its performance significantly surpasses all fully fine-tuned baselines. This phenomenon led us to propose a core hypothesis: the performance advantage of the TTPrompt framework is positively correlated with the extraction complexity of the dataset. To quantitatively verify this hypothesis, we designed the correlation analysis experiment in this section.

For this analysis, we first introduce two key metrics. The first is the dataset difficulty index $\Omega$, which quantifies the intrinsic complexity of a dataset across six dimensions, with its detailed calculation method described in Appendix~\ref{sec:Difficulty} and Table~\ref{tab:difficulty_index_rounded}. The second is the performance gain $\Delta F1$, defined as $\Delta F1 = F1_{TTPrompt} - F1_{\text{Fine-tuned}}$. For the Fine-tuned baseline, we chose LADDER*, a mainstream method in CTI NER, while for the CTINexus dataset, we used the results from ACLM.

\begin{figure}[!h]
    \centering
    \includegraphics[width=0.75\linewidth]{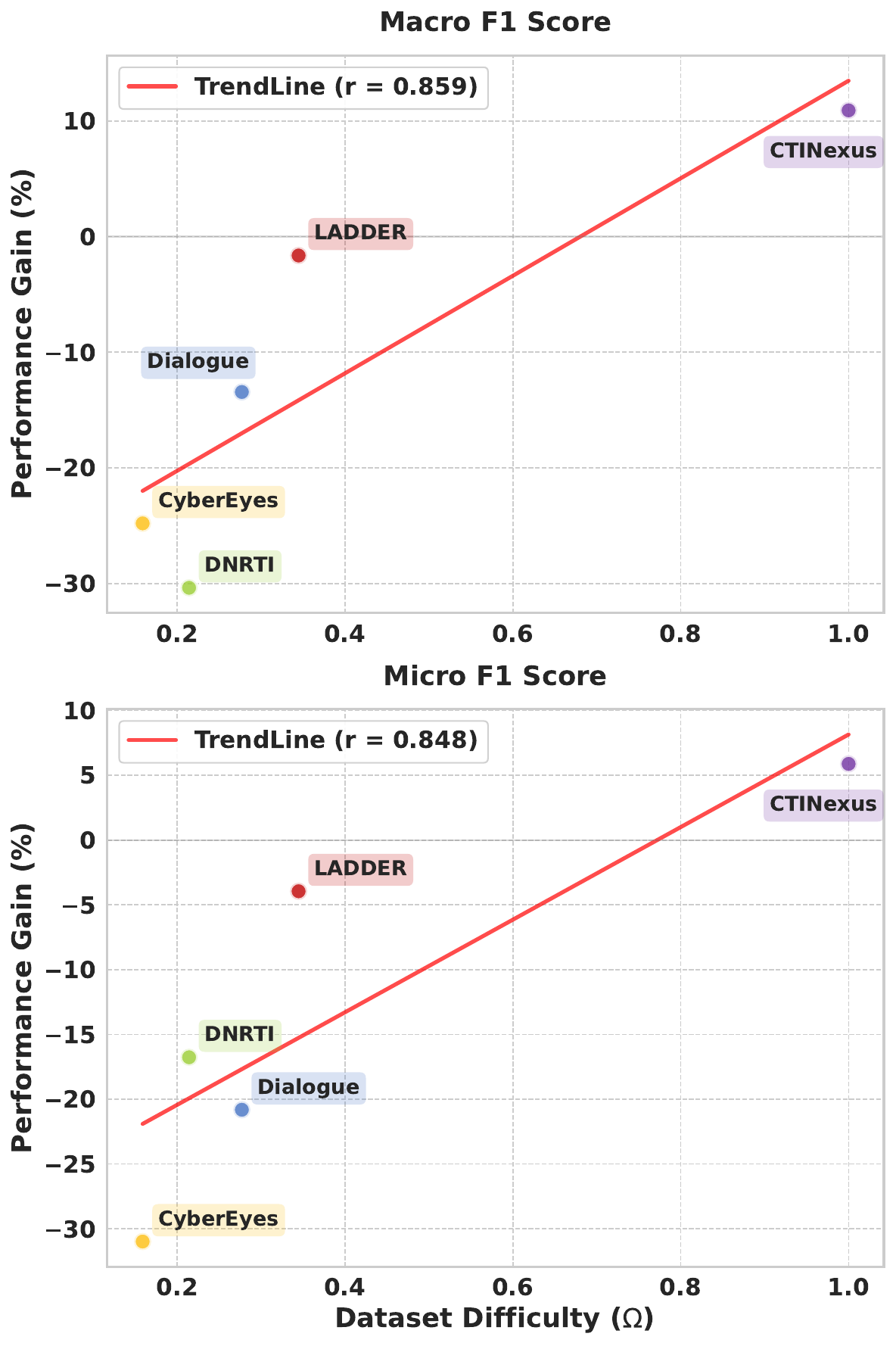}
    \caption{Performance Gain ($\Delta F1$) vs. Dataset Difficulty ($\Omega$).}
    \label{fig:Dataset_Omega}
\end{figure}

As shown in Figure~\ref{fig:Dataset_Omega}, the graph clearly reveals a strong positive correlation between the two. On datasets with a low difficulty index, such as CyberEyes with an $\Omega$ value of 15.89\%, the $\Delta F1$ is negative, indicating that fully fine-tuned models still have an advantage on such tasks. However, as the $\Omega$ value increases, TTPrompt's relative performance monotonically improves, reaching its maximum positive value on the most difficult dataset, CTINexus, where $\Omega$ is 1. This observation perfectly aligns with our findings in the main experiments. 
The r-values for Macro F1 and Micro F1 reached 0.859 (p=0.0625) and 0.848 (p=0.0697), respectively. While the p-values are slightly above the 0.05 significance threshold due to the small sample size (N=5), the strong correlation of approximately 0.85 still offers powerful support for our hypothesis.\looseness=-1

To further contextualize our method's data efficiency, we also evaluated the performance of fine-tuning LADDER* and ACLM in a similar low-resource setting to investigate the feasibility of traditional fine-tuning paradigms in this scenario. The results reveal the extreme difficulty of low-resource CTI NER tasks. As shown in Table~\ref{tab:low_sft}, a standard fine-tuned model like LADDER* scores an F1 of 0 on most datasets when trained on a small number of samples, indicating a complete inability to learn CTI knowledge in a low-resource setting. ACLM, a model specifically designed for low-resource NER, is particularly noteworthy. Even so, it only achieves effective non-zero results on the datasets with the lowest difficulty index $\Omega$, such as CyberEyes and CyberDialogue. Even on CyberDialogue, where it performed relatively well, its Macro F1 was only 10.57\%. In contrast, TTPrompt, which also undergoes instruction refinement in a 1\% low-resource setting, achieves a Macro F1 of 46.92\%. This vast gap demonstrates that our method of optimizing instructions based on limited feedback exhibits far superior performance compared to traditional paradigms.

\subsection{RQ4: Effect of Strategy Generation Models}

To validate the effectiveness of the guiding strategy component and to select the best base model for the strategy generation agent, we conducted a comparative experiment. We used three of the most representative current models, DeepSeek-v3.1~\cite{guo2025deepseek}, GPT-5~\footnote{https://chatgpt.com/}, and Gemini 2.5 Pro~\cite{comanici2025gemini}, to each generate 10 different guiding strategies. These 30 strategies were independently evaluated on a 1\% training subset of four core datasets. All performance was measured by the Micro F1 score.

\begin{table*}[!h]
\setlength{\abovecaptionskip}{0.1cm}
\setlength{\belowcaptionskip}{-0.2cm}
    \centering
    \caption{Comparison of teacher model guided strategies on the LADDER and CTINexus datasets.}
    \label{tab:guidance_merged}
    \begin{tabular}{l|cc|cc||cc|cc}
        \hline
        \multirow{3}{*}{\textbf{Strategy}} & \multicolumn{4}{c||}{\textbf{LADDER}} & \multicolumn{4}{c}{\textbf{CTINexus}} \\
        \cline{2-9}
        ~ & \multicolumn{2}{c|}{\textbf{Macro}} & \multicolumn{2}{c||}{\textbf{Micro}} & \multicolumn{2}{c|}{\textbf{Macro}} & \multicolumn{2}{c}{\textbf{Micro}} \\
        \cline{2-9}
        ~ & same & qwen & same & qwen & same & qwen & same & qwen \\
        \hline \hline
        DS-C-3.1 & 65.60\% & \textcolor{myorange}{64.25\%} & 69.82\% & 69.04\% & 19.94\% & \textcolor{myorange}{21.82\%} & 30.23\% & \textcolor{myorange}{31.75\%} \\
        DS-R-3.1 & \textcolor{myorange}{68.05\%} & 64.24\% & \textcolor{myorange}{72.32\%} & \textcolor{myorange}{69.70\%} & 18.46\% & 20.46\% & \textcolor{myorange}{34.48\%} & 30.86\% \\
        DS-C-3.2 & 64.52\% & \textcolor{myred}{65.59\%} & 69.43\% & \textcolor{myred}{70.92\%} & \textcolor{myred}{23.18\%} & \textcolor{myred}{23.58\%} & 29.44\% & \textcolor{myred}{32.38\%} \\
        DS-R-3.2 & \textcolor{myred}{68.44\%} & 63.52\% & \textcolor{myred}{72.76\%} & 69.34\% & \textcolor{myorange}{21.36\%} & 21.63\% & 34.04\% & 32.63\% \\
        \hline
    \end{tabular}
\end{table*}

The experimental results, as shown in Figure~\ref{fig:Strategies}, clearly reveal the superiority of DeepSeek-v3.1 in the strategy generation task. Across all four datasets, the strategies generated by DeepSeek-v3.1 generally exhibited the strongest and most stable performance, and the upper bound of their best strategy's performance was typically the highest. Although Gemini 2.5 Pro was also able to generate some high-performance strategies, its overall stability was inferior to that of DeepSeek-v3.1, while GPT-5's performance was relatively weaker. It is worth noting that a huge variance exists in the performance of strategies from all models, which strongly justifies the necessity of the Strategy Selection step in our methodology.

\subsection{RQ5: Cross-Model Transferability of Guidelines}

The experiment in this section aims to investigate the transferability of the annotation guidelines in our methodology; that is, whether an annotation guideline generated by a high-capability teacher model can be effectively executed by a student model. To answer this question, we designed a cross-model transfer effectiveness experiment.

\textbf{Experimental Setup}
For the experimental setup, we selected different versions of the DeepSeek (DS) model, including the DS-chat (DS-C) and the DS-Reasoner (DS-R)  versions, to be responsible for generating the guiding strategies. Meanwhile, we chose the Qwen3-32B (Qwen) model as the student model, responsible for executing the NER task. We designed two experimental conditions and evaluated them on the LADDER and CTINEXUS datasets: (1) Same-Model baseline: Both the generation of the annotation guideline and the execution of the task were performed by the same DS model. (2) Cross-Model transfer: The annotation guideline was generated by a DS model but was executed by the Qwen model.\looseness=-1

\textbf{Results Analysis}
As shown in Table~\ref{tab:guidance_merged}, the experimental results clearly demonstrate the excellent portability of the annotation guidelines. In all experiments, the strategies generated by the DS models were able to successfully guide the Qwen model to complete the task while maintaining a considerably high level of performance. For example, on the LADDER dataset, when the strategy generated by DS-R-3.2 was transferred to the Qwen model for execution, the Micro F1 score only slightly decreased from 72.76\% to 69.34\%, proving the effective transfer of the strategy.

\subsection{RQ6: Analysis of Epochs and Data Volume}
To deeply investigate the characteristics of the FIR mechanism, we analyzed its two key hyperparameters: training epochs and the amount of training data. The experiments in this section aim to answer RQ6.

\begin{figure}[h]
    \centering
    \includegraphics[width=0.85\linewidth]{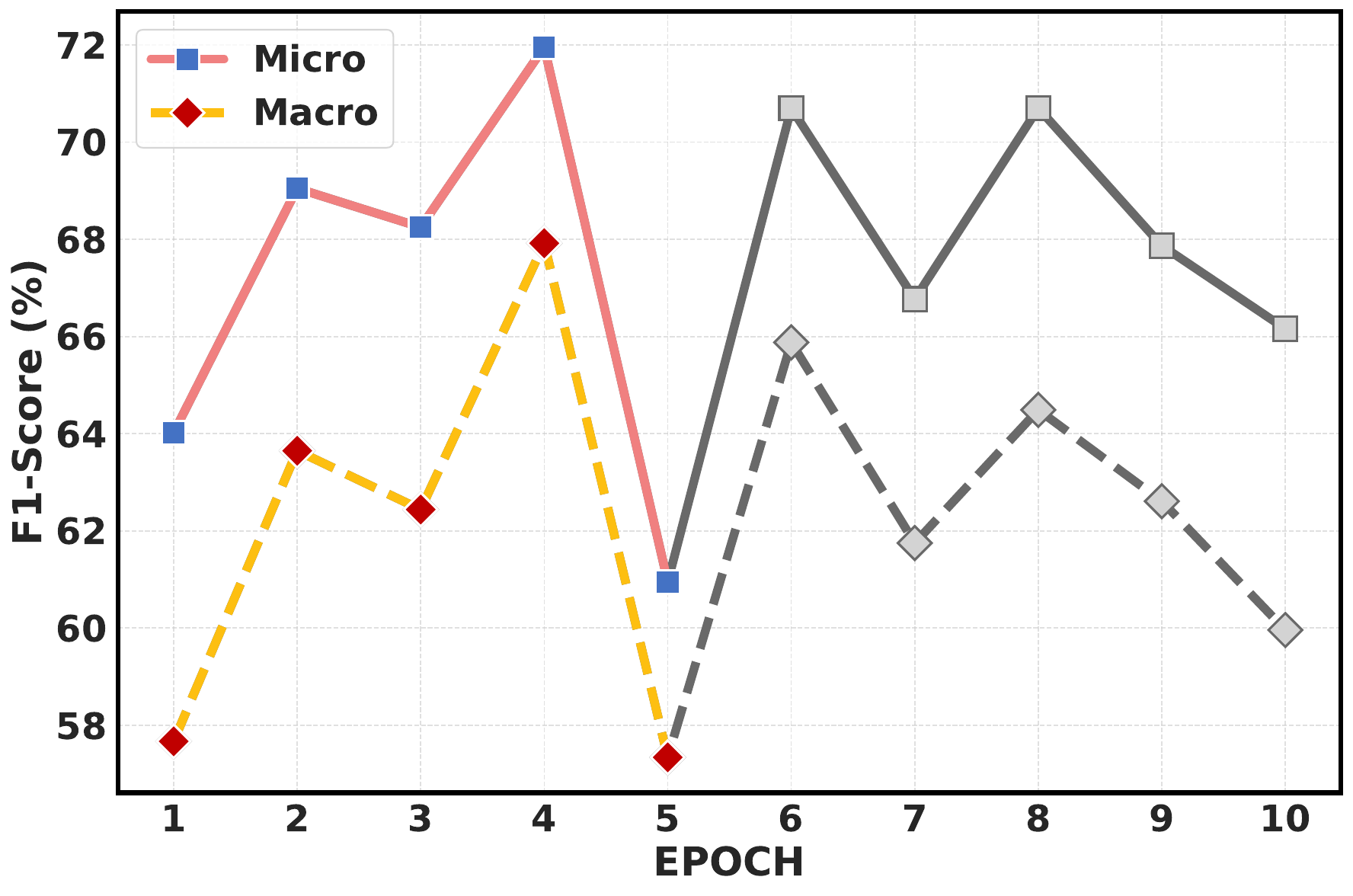}
    \caption{Performance of the FIR process over 10 epochs.}
    \label{fig:EPOCH}
\end{figure}

We first iterated the FIR process for 10 consecutive epochs on a 1\% subset of the training data, recording the model's performance on the validation set after each epoch. As shown in Figure~\ref{fig:EPOCH}, the model's Micro F1 and Macro F1 scores peaked at the 4th epoch before starting to fluctuate or even decline. The performance curve did not continue to monotonically increase with more training epochs. This phenomenon indicates that the FIR process converges very quickly. The initial few epochs are sufficient for the LLM to effectively correct the critical flaws in the instruction set based on the most significant error signals. However, conducting too many rounds of iteration on the same small-scale dataset may cause the instructions to overfit to the samples of that specific subset. This could introduce revisions that reduce generalization ability and negatively impact overall performance. Therefore, we determined that a large number of training epochs is unnecessary; a small number, such as the 5 epochs used in our experiments, is both sufficient and efficient.


\begin{table}[h]
\setlength{\abovecaptionskip}{0.1cm}
\setlength{\belowcaptionskip}{-0.2cm}
    \centering
    \caption{Effect of training data size.}
    \label{tab:training_percentage}
    \begin{tabular}{c|cc}
        \hline
        \textbf{Training Set (\%)} & \textbf{Macro} & \textbf{Micro} \\
        \hline \hline
        0.5 & \textcolor{myorange}{66.46\%} & 71.38\% \\
        1   & \textcolor{myred}{67.92\%} & \textcolor{myred}{71.96\%} \\
        2   & 58.94\% & 69.53\% \\
        3   & 62.56\% & \textcolor{myorange}{71.59\%} \\
        \hline
    \end{tabular}
\end{table}

Next, we investigated the impact of the amount of data used for instruction refinement on the final performance. We used 0.5\%, 1\%, 2\%, and 3\% of the training data, respectively, to perform instruction refinement for a fixed period. The results are shown in Table~\ref{tab:training_percentage}. The experimental data indicates that model performance is optimal when the data volume is at 1\%. It is noteworthy that with only 0.5\% of the data, the performance is already very close to the optimal level. However, when we further increased the data volume to 2\% and 3\%, the model's performance showed a noticeable decline.

This result strongly demonstrates the extremely high data efficiency of the FIR mechanism. We speculate that the reason for the performance decrease may be that as more data is introduced, the generated semantic gradients become more complex and diverse, possibly even containing conflicting revision suggestions. This increases the difficulty for the LLM to learn a general, high-quality instruction set from a single correction. In contrast, a small yet representative dataset (such as 0.5\%-1\%) can provide more consistent improvement signals.\looseness=-1

\section{Conclusion}



This paper presents an in-depth empirical analysis of the mainstream RetICL paradigm on CTI NER tasks. Our study uncovers a fundamental flaw: its performance arises not from global semantic understanding but from incidental entity-type overlap in retrieved examples, revealing the unreliability of implicit induction and the limits of learning from unstable cases.\looseness=-1

To overcome the limitations of implicit induction, we propose TTPrompt, a paradigm shifting from implicit induction to explicit instruction. Structurally, TTPrompt constructs a deductive reasoning pathway that mirrors expert cognition, mapping the CTI analysis process into a hierarchical framework of TTPs. Furthermore, to resolve the adaptability bottleneck of static prompts, we introduce the FIR mechanism. Functioning as a semantic optimizer, FIR mimics the backpropagation algorithm by calculating semantic gradients from prediction errors, enabling the model to self-correct and iteratively refine its guidelines using only 1\% of the training data. Extensive experiments on five CTI benchmarks demonstrate that TTPrompt not only consistently surpasses RetICL but, without parameter tuning, achieves performance comparable to fully fine-tuned models, and even exceeds them on complex datasets, offering a more efficient NER solution in complex domains.

\section{ACKNOWLEDGMENT}
This work was supported by National Natural Science Foundation of China (No.62472296), Sichuan Science and Technology Program (No.2024ZHCG0197).

We thank the anonymous reviewers for their valuable feedback and insightful comments, which significantly helped us
enhance the quality and clarity of our manuscript.

\bibliographystyle{IEEEtran}
\bibliography{medds}

\appendix

\subsection{Datasets}
\label{sec:Datasets}
For a comprehensive evaluation, we selected five uniquely challenging CTI NER datasets. These datasets are distinct in terms of language, text genre, and classification difficulty.

\begin{table}[!htbp]
\setlength{\abovecaptionskip}{0.1cm}
\setlength{\belowcaptionskip}{-0.2cm}
    \centering
    \caption{Statistics of the datasets used in our experiments.}
    \label{tab:dataset_statistics}
    \begin{tabular}{l|l|c|l}
        \hline
        \textbf{Dataset} & \textbf{Language} & \textbf{Entity Types} & \textbf{Train/Dev/Test} \\
        \hline \hline
        LADDER        & English & 9  & 3666/749/736 \\
        CTINexus      & English & 24 & 137/---/10 \\
        CyberDialogue & Chinese & 5  & 3283/---/365 \\
        CyberEyes     & English & 4  & 15236/2217/1808 \\
        DNRTI         & English & 13 & 5251/662/664 \\
        \hline
    \end{tabular}
\end{table}

\textbf{LADDER:} One of the most widely used datasets in the CTI NER domain.

\textbf{CTINexus:} This dataset is extremely challenging, primarily due to: 1) long input texts, 2) a large number of entity types, 3) fine-grained category definitions, and 4) long entity spans. We used the data splits from the original paper to partition all datasets and removed two duplicate samples.

\textbf{CyberDialogue:} This is a dataset in a Chinese conversational format. Its inclusion aims to test the method's generalization ability across different languages and text genres.

\textbf{CyberEyes:} In contrast to CTINexus, CyberEyes has a relatively small number of entity types. 

\textbf{DNRTI:} The notable feature of this dataset is its extremely high entity density. Only 1.18\% of the samples in its training set contain no entities, and this proportion is as low as 0.30\% in the test set.

Following previous work~\cite{ma2023large}, and given the resource limitations associated with API access, we sampled approximately one-third of the original test data for each benchmark. This resulted in 200 instances for DNRTI and 600 instances for CyberEyes.
For all datasets, unless otherwise specified, FIR uses 1\% of the training set (5\% for CTINexus, i.e., 6 samples, due to its small size). Similar to~\cite{luo2025oneke}, we employ iterative stratified sampling for the train sets. LADDER and ACLM use the full training set. For datasets without a validation set, we follow~\cite{he2023revisiting} and use the training set for validation and as the demonstration pool. For those with a validation set, CTINexus* uses its validation set as the demonstration pool. For more statistical data, see Table~\ref{tab:dataset_statistics}.


\subsection{Dataset Difficulty Index}
\label{sec:Difficulty}

\begin{table*}[!h]
\setlength{\abovecaptionskip}{0.1cm}
\setlength{\belowcaptionskip}{-0.2cm}
    \centering
    \caption{Analysis of dataset characteristics and calculation of the Difficulty Index ($\Omega$).}
    \label{tab:difficulty_index_rounded}
    \begin{tabular}{l|cccccc|c}
       \hline
          \textbf{Dataset} & \textbf{$L_{doc}$} & \textbf{$N_{type}$} & \textbf{$S_{type}$} & \textbf{$L_{ent}$} & \textbf{$C_{type}$} & \textbf{$N_{ent}$} & \textbf{$\Omega$} \\
          \hline \hline
       CTINexus    & 1.00 & 1.00 & 1.00 & 1.00 & 1.00 & 1.00 & \textbf{1.00} \\
       LADDER    & 0.08 & 0.25 & 0.31 & 0.22 & 0.61 & 0.60 & \textbf{0.34} \\
       CyberDialogue    & 0.00 & 0.05 & 0.91 & 0.00 & 0.56 & 0.15 & \textbf{0.28} \\
       DNRTI   & 0.08 & 0.45 & 0.00 & 0.40 & 0.35 & 0.00 & \textbf{0.21} \\
       CyberEyes & 0.06 & 0.00 & 0.27 & 0.22 & 0.00 & 0.40 & \textbf{0.16} \\
       \hline
    \end{tabular}
\end{table*}

To quantitatively analyze the complexity of different datasets, we have designed a comprehensive Dataset Difficulty Index, $\Omega$. This index aims to quantify the intrinsic challenges of an NER task from multiple dimensions.

\textbf{Average Document Length ($L_{doc}$):} Tests the model's ability to handle long-range dependencies and to locate sparse information within a broad context.

\textbf{Number of Entity Types ($N_{type}$):} Directly reflects the scale of the class boundaries that the model needs to learn and distinguish.

\textbf{Entity Type Sparsity ($S_{type}$):} Measures the imbalance in the distribution of entity types. A highly imbalanced long-tail distribution significantly increases learning difficulty. We use the Gini Coefficient to quantify this metric.

\textbf{Average Entity Length ($L_{ent}$):} Refers to the average length of entity text spans, testing the model's ability to perform precise boundary detection.

\textbf{Entity Type Confusability ($C_{type}$):} Measures the proximity of different entity classes in the semantic space. A higher value indicates that the boundaries between entity classes are more ambiguous, demanding a higher fine-grained discrimination ability from the model. We use all-MiniLM-L6-V2~\footnote{https://huggingface.co/sentence-transformers/all-MiniLM-L6-v2} for embeddings.

\textbf{Entity Novelty ($N_{ent}$):} This dimension measures the lexical novelty of the test set relative to the training set. It quantifies the proportion of entities that appear in the test set but were never seen in the training set.

We perform Min-Max Normalization on the quantified values corresponding to the six dimensions mentioned above:
\begin{equation}
    X_i \in \{L_{\text{doc}}, N_{\text{type}}, S_{\text{type}}, L_{\text{ent}}, C_{\text{type}}, N_{\text{ent}}\}
\end{equation}
Subsequently, we scale these values to the $[0, 1]$ range to eliminate the effects of different scales.
The $\Omega$ is calculated by a weighted sum of the normalized indicators for each dimension:
\begin{equation}
\Omega(D) = \sum_{i=1}^{6} w_i \cdot \overline{X_i}
\end{equation}
where $\overline{X_i}$ represents the normalized value of the $i$-th dimension, and $w_i$ is its corresponding weight (satisfying $\sum w_i = 1$). In this study, we assign equal values to all weights, i.e., $w_i = 1/6$.


\subsection{Details of Baselines}
\label{sec:baseline_details}

\textbf{LADDER*:} This framework is a representative work in the CTI domain that performs NER by fine-tuning a Pre-trained Language Model (PLM). We chose LADDER* as a baseline to measure our method's performance against the mainstream, traditional fine-tuning paradigm in the field.

\textbf{ACLM:} This is a data augmentation framework specifically designed for low-resource and complex NER tasks. It is based on a PLM and generates high-quality augmented data through an innovative selective denoising method. ACLM is included to test the effectiveness of our method in handling challenging scenarios with data sparsity and complex entities.

\textbf{OneKE:} This system is a multi-agent knowledge extraction framework based on an LLM. It performs extraction through a collaborative workflow of agents with schema, extraction, and reflection roles, utilizing a Case Repository for multi-turn dialogue. We chose OneKE to directly compare our method with contemporary LLM systems that employ multi-agent collaboration paradigms.

\textbf{CTINEXUS*:} This framework is a CTI knowledge extraction framework based on an LLM. Its core mechanism is to use a kNN retriever to dynamically select the most relevant examples for each input, thereby constructing a prompt to guide the LLM's ICL. We have modified it to focus on the NER task, with details available in Appendix~\ref{sec:kNN-based}.

\subsection{Implementation Details}
\label{sec:Implementation}
Unless otherwise specified, all experiments in this paper involving LLMs use Qwen3-32B~\cite{yang2025qwen3}. We use the vLLM v0.9.2~\cite{kwon2023vllm} framework for local deployment and inference acceleration and the openai v1.93.0 library for API calls. The temperature was set to 0. In the FIR framework, the number of epochs was 5, the batch size was 1. All experimental procedures, including the reproduction of baseline models, were completed on 2 * NVIDIA A800 80GB GPUs.

\subsection{kNN-based ICL Paradigm}
\label{sec:kNN-based}
\subsubsection{\textbf{kNN-based Demonstration Retriever}}
The objective of this component is to retrieve the k most relevant demonstration examples for a query document $W_{q}$ (where $W_{q} \in D_{s}$) from an existing annotated dataset $D_{u}$. In the original work and this paper, k=3. Each sample in $D_{u}$ can be represented as a pair $(W^{(j)}, \mathcal{E}^{(j)})$, where $W^{(j)}$ is the token sequence of the document and $\mathcal{E}^{(j)}$ is its corresponding ground-truth entity set.

The retrieval process is formalized as follows: First, a pre-trained text embedding model $\phi(\cdot)$, such as text-embedding-3-large, is used to compute high-dimensional vector representations for the query document $W_{q}$ and all candidate documents $W^{(j)}$ in $D_{u}$:
\begin{equation}
\begin{gathered}
        v_{q} = \phi(W_{q}) \\
        v_{j} = \phi(W^{(j)}), \forall W^{(j)} \in D_{u}
\end{gathered}
\label{eq:equation2}
\end{equation}

Next, the cosine similarity $s_{j}$ is calculated between the query vector $v_{q}$ and each candidate document vector $v_{j}$:
\begin{equation}
s_{j} = \frac{v_{q} \cdot v_{j}}{\|v_{q}\| \|v_{j}\|}
\end{equation}
Finally, all candidate documents are ranked based on the similarity score $s_{j}$, and the k samples with the highest scores are selected as the demonstration set $\mathcal{D}_{k}$. This process can be represented as:
\begin{equation}
\mathcal{D}_{k} = \text{Top-k}_{j \in D_{u}} \{ (W^{(j)}, \mathcal{E}^{(j)}) | s_{j} \}
\end{equation}

\subsubsection{I\textbf{CL Prompt Construction and Inference}}
\label{sec:ICL Prompt Construction and Inference}
After obtaining the demonstration set $\mathcal{D}_{k}$, the method constructs a structured ICL prompt $\mathcal{P}$. This prompt $\mathcal{P}$ consists of three parts:

\textbf{Task Instruction ($\mathcal{T}_{\text{instr}}$):} Clearly defines the objective of the NER task, the set of entity types Y to be identified, and the desired output format.

\textbf{Demonstration Examples ($\mathcal{D}_{k}$):} The k examples obtained from the retriever. These examples are not randomly ordered but are sorted in ascending order of their similarity score $s_{j}$ with the query document. This leverages the LLM's Recency Bias, placing the example most similar to the query closest to it. If the examples in $\mathcal{D}_{k}$ are sorted by similarity in ascending order into a sequence $\mathcal{S}_{k} = (d_1, d_2, ..., d_k)$, then: $s_{d_1} \le s_{d_2} \le ... \le s_{d_k}$.

\textbf{Query Document ($W_{q}$):} The original input text for which NER is to be performed.

The final structure of the prompt $\mathcal{P}$ is constructed by concatenating the task instruction, the retrieved demonstrations, and the query document, formalized as follows:
\begin{equation}
    \mathcal{P} = [\mathcal{T}_{\text{instr}}; d_1; d_2; \dots; d_k; W_{q}]
\end{equation}
This fully constructed prompt $\mathcal{P}$ is submitted as a single query to the LLM. The model directly generates the set of entities $\mathcal{E}_{q}$ contained in the query document $W_{q}$, represented as:
\begin{equation}
    \mathcal{E}_{q} = \text{LLM}(\mathcal{P})
\end{equation}

\subsection{Type-Overlap Paradigm (Oracle)}
\label{sec:Type-Overlap}
The Type-Overlap paradigm is designed as an Oracle experiment to quantify the theoretical impact of entity type overlap. It operates on the idealized premise of knowing the ground-truth entity types of the query document $W_q$ in advance. Its retrieval objective is to select demonstration examples that share the highest number of entity types with the query document.

\textbf{Demonstration Retrieval.}
The retrieval process is formalized as follows:
First, we define a function $\text{typeset}(\mathcal{E})$ that extracts the set of unique entity types from a given entity set $\mathcal{E}$:
\begin{equation}
\text{typeset}(\mathcal{E}) = \{ y_m | (\text{span}, y_m) \in \mathcal{E} \}
\end{equation}
For a query document $W_q$, its ground-truth entity set is represented as $\mathcal{E}_q$. The Oracle information utilized by this paradigm is its true set of entity types, $\text{typeset}(\mathcal{E}_q)$. For each candidate sample $(W^{(j)}, \mathcal{E}^{(j)})$ in the retrieval pool $D_{u}$, we calculate its entity type overlap score $o_j$ with the query document. This score is the size of the intersection of their respective entity type sets:
\begin{equation}
o_j = | \text{typeset}(\mathcal{E}_q) \cap \text{typeset}(\mathcal{E}^{(j)}) |
\end{equation}
Finally, the k samples with the highest overlap scores $o_j$ are selected to form the demonstration set $\mathcal{D}_k$:
\begin{equation}
\mathcal{D}_k = \text{Top-k}_{j \in D_{u}} \{ (W^{(j)}, \mathcal{E}^{(j)}) | o_j \}
\end{equation}

The prompt construction process is the same as in the Section~\ref{sec:ICL Prompt Construction and Inference}.

\subsection{Entity-Density Paradigm}
\label{sec:Entity-Density}
The Entity-Density paradigm is a simple retrieval paradigm designed to test the model's performance when faced with examples of extremely high information density. This method completely ignores the content of the query document $W_q$ and selects candidates based solely on the number of annotated entities they contain.

\textbf{Demonstration Retrieval.}
The retrieval process is formalized as follows:
For each candidate sample $(W^{(j)}, \mathcal{E}^{(j)})$ in the retrieval pool $D_{u}$, we calculate its entity density score $c_j$, defined as the total number of entity instances in that sample: $c_j = |\mathcal{E}^{(j)}|$.
The k samples with the highest density scores $c_j$ are selected to form the demonstration set $\mathcal{D}_k$:
\begin{equation}
\mathcal{D}_k = \text{Top-k}_{j \in D_{u}} \{ (W^{(j)}, \mathcal{E}^{(j)}) | c_j \}
\end{equation}

The prompt construction process is the same as in the Section~\ref{sec:ICL Prompt Construction and Inference}.

\begin{table*}[h]
\centering
\caption{Case Study of FIR's Iterative Guideline Refinement from LADDER dataset.}
\label{tab:case_study}
\begin{tabular*}{\textwidth}{@{\extracolsep{\fill}}ll}
\toprule

\multicolumn{2}{l}{\cellcolor{lightgraybg}\textbf{Epoch 1 \(\rightarrow\) 2: Correcting Over-generalization}} \\
\midrule
\textbf{Case Text} &
\parbox[t]{0.8\textwidth}{
    \textit{Using these permissions, the malware can send messages, make calls
    and perform \textcolor{myred}{\textbf{overlay attacks}} easily without \\ alerting the user.}
} \\

\addlinespace[5pt]
\textbf{Guideline Evolution} &
\parbox[t]{0.8\textwidth}{
\textcolor{StepBlue}{\textbf{Initial Guideline (Epoch 1):}}
\begin{itemize}
    \item[\textbullet] \textbf{Malware Type:} Generic terms describing functional categories or behaviors. Exclude basic technical concepts like 'backdoor', 'Trojan', and functional descriptive terms like 'information stealer'.
\end{itemize}
\vspace{-1ex}
\textcolor{StepGold}{\textbf{Updated Guideline (Epoch 2):}}
\begin{itemize}
    \item[\textbullet] \textbf{Malware Type:} \textcolor{myorange}{\textbf{Label only when a term in the context explicitly refers to an independently classifiable malware behavior pattern (e.g., 'Ransomware', 'Worm').}} Exclude generic terms like 'malware'. When such terms act as modifiers (e.g., in 'Rotexy Trojan'), they should be labeled separately.
\end{itemize}
} \\

\addlinespace[5pt]
\textbf{Prediction Evolution} &
\parbox[t]{0.8\textwidth}{
\textbf{Epoch 1 Prediction:} \{\texttt{Malware Type}: [``overlay attacks'']\} \(\rightarrow\) \textcolor{myred}{\textbf{\textit{False Positive}}} \\
\textbf{Epoch 2 Prediction:} \{\} \(\rightarrow\) \textcolor{StepBlue}{\textbf{\textit{Correction Successful}}}
} \\
\addlinespace[5pt]
\textbf{Analysis} &
\cellcolor{MyBoxBg}
\parbox[t]{0.8\textwidth}{
    The initial broad guideline led to misclassifying an attack method. A single FIR iteration successfully tightened the semantic boundary by adding a negative constraint, correcting the false positive.
} \\
\midrule

\multicolumn{2}{l}{\cellcolor{lightgraybg}\textbf{Epoch 2 \(\rightarrow\) 3: Learning Structural Rules}} \\
\midrule
\textbf{Case Text} & \textit{\textcolor{myred}{\textbf{DroidJack RAT}} starts capturing sensitive information like call data , SMS data , videos , photos , etc.} \\

\addlinespace[5pt]
\textbf{Guideline Evolution} &
\parbox[t]{0.8\textwidth}{
\textcolor{StepBlue}{\textbf{Guideline Before (Epoch 2):}}
\begin{itemize}
    \item[\textbullet] \textbf{Malware:} Specific malware names. If a term is both a type and a family name, prioritize Malware Type.
    \item[\textbullet] \textbf{Malware Type:} Generic terms referring to independently classifiable malware behavior patterns.
\end{itemize}
\vspace{-1ex}
\textcolor{StepGold}{\textbf{Updated Guideline (Epoch 3):}}
\begin{itemize}
    \item[\textbullet] \textbf{Malware:} \textcolor{myorange}{\textbf{When a name is followed by a standardized type term, it should be split into separate Malware and Malware Type entities.}} When a name includes modifiers (e.g., 'Mobile Rotexy Malware'), label only the core proper noun 'Rotexy'.
    \item[\textbullet] \textbf{Malware Type:} Label only when a term in the context explicitly refers to a \textbf{standardized type} within a known malware classification system (e.g., 'Ransomware', 'Trojan').
\end{itemize}
} \\

\addlinespace[5pt]
\textbf{Prediction Evolution} &
\parbox[t]{0.8\textwidth}{
\textbf{Epoch 1/2 Prediction:} \{\texttt{Malware}: [``DroidJack RAT'']\} \(\rightarrow\) \textcolor{myred}{\textbf{\textit{Boundary Error}}} \\
\textbf{Epoch 3 Prediction:} \{\texttt{Malware}: [``DroidJack''], \texttt{Malware Type}: [``RAT'']\} \(\rightarrow\) \textcolor{StepBlue}{\textbf{\textit{Correction Successful}}}
} \\
\addlinespace[5pt]
\textbf{Analysis} &
\cellcolor{MyBoxBg}
\parbox[t]{0.8\textwidth}{
FIR induced a new structural rule from accumulated boundary errors, enabling the model to parse composite entities correctly instead of treating them as a single token.
} \\
\midrule

\multicolumn{2}{l}{\cellcolor{lightgraybg}\textbf{Epoch 3 \(\rightarrow\) 4: Refining Semantic Boundaries}} \\
\midrule
\textbf{Case Text} & \parbox[t]{0.8\textwidth}{
    \textit{For contacting C \& C, the \textcolor{myred}{\textbf{spyware}} was found to be using free DNS services , as shown in the screenshot below: \textcolor{myred}{\textbf{SpyNote RAT}} uses an unusual trick to make sure that it remains up and running and that the spying does not stop.}
} \\

\addlinespace[5pt]
\textbf{Guideline Evolution} &
\parbox[t]{0.8\textwidth}{
\textcolor{StepBlue}{\textbf{Guideline Before (Epoch 3):}}
\begin{itemize}
    \item[\textbullet] \textbf{Malware Type:} Refers to standardized types in known classification systems.
\end{itemize}
\vspace{-1ex}
\textcolor{StepGold}{\textbf{Updated Guideline (Epoch 4):}}
\begin{itemize}
   \item[\textbullet] \textbf{Malware Type:} \textbf{Specific technical terms} describing functional categories (e.g., bot, ransomware, botnet). \textcolor{myorange}{\textbf{Exclude terms for attack methods or exploits (e.g., exploits, phishing, backdoor)}}, and exclude generic terms like 'malware', 'information stealer'.
\end{itemize}
} \\
\addlinespace[5pt]
\textbf{Prediction Evolution} &
\parbox[t]{0.8\textwidth}{
\textbf{Epoch 1-3 Prediction:} \{\texttt{Malware}: [``SpyNote RAT''], \texttt{Malware Type}: [``RAT'']\} \(\rightarrow\) \textcolor{myred}{\textbf{\textit{Classification Error}}} \\
\textbf{Epoch 4 Prediction:} \{\texttt{Malware}: [``SpyNote RAT''], \texttt{Malware Type}: [``spyware'']\} \(\rightarrow\) \textcolor{StepBlue}{\textbf{\textit{Correction Successful}}}
} \\
\addlinespace[5pt]
\textbf{Analysis} &
\cellcolor{MyBoxBg}
\parbox[t]{0.8\textwidth}{
After multiple iterations, FIR learned to encode the dataset's annotation dialect, making more contextually appropriate semantic choices (e.g., preferring "spyware" from context over "RAT").
} \\
\bottomrule
\end{tabular*}
\end{table*}

\end{document}